\documentclass[prl,twocolumn,showpacs]{revtex4}
\usepackage{graphicx}

\begin{document}

\title{The Equilibrium Intrinsic Crystal-Liquid Interface of Colloids}

\author{Jessica Hern{\'a}ndez-Guzm{\'a}n}
\author{Eric R.~Weeks}
\email{weeks@physics.emory.edu}
\affiliation{ Department of Physics, Emory University, Atlanta,
GA
30322\\
Submitted to Proceedings of the National Academy of
Sciences of the United States of America}

\date{\today}


\begin{abstract}
We use confocal microscopy to study an equilibrated crystal-liquid
interface in a colloidal suspension.  Capillary waves roughen the
surface, but locally the intrinsic interface is sharply defined.
We use local measurements of the structure and dynamics to
characterize the intrinsic interface, and different measurements
find slightly different widths of this interface.  In terms
of the particle diameter $d$, this width is either $1.5d$ (based
on structural information) or $2.4d$ (based on dynamics), both
not much larger than the particle size.  This work is the first
direct experimental visualization of an equilibrated crystal-liquid
interface.
\end{abstract}

\maketitle

\section{Introduction}

The interface between crystal and liquid phases of a
material governs phenomena such as wetting, lubrication, and
crystal nucleation \cite{kaplan06,laird98}.  Interfaces
are poorly defined at the atomic level: capillary
waves cause fluctuations in the interface position
\cite{stillinger65,muller05,aarts04,kegel06}, and
locally the structure varies in a smooth way from ordered
to disordered \cite{laird98}.  The literature makes a
distinction between the intrinsic interface (presumed to be
sharp) \cite{stillinger65}, and the observed surface blurred
by capillary waves \cite{muller05,chacon06}.  The standard
equilibrium interface profile is well-defined and contains
fluctuations at all length and time scales, due to these
capillary waves.  Because of
the rapid
time scales of capillary wave fluctuations, along with the small
length scales at the interface, it is difficult to study these
interfaces directly \cite{kaplan06}.  Thus computer simulations
provide useful information about model crystal/liquid interfaces,
such as hard sphere systems \cite{laird98,laird05b,laird05} and
Lennard-Jones systems \cite{broughton86,huitema99}.

Recently, crystal/liquid interfaces were directly
studied in colloidal suspensions using confocal microscopy
\cite{kegel06,gasser01}.  Colloids are systems of solid particles
in a liquid, and are a good model system for phase transitions
\cite{aarts04,kegel06,pusey86,ackerson95}.  Microscopy allows direct
observation of structure and dynamics of the colloidal particles
\cite{prasad07}.  However, the previous experiments focused on
nonequilibrium cases where samples were crystallizing, and did not
provide data on equilibrium interfaces such as those studied by
simulation \cite{laird98,laird05b,laird05,broughton86,hoyt01}.
Furthermore, these experiments did not examine the intrinsic
interface, perhaps because they were nonequilibrium studies and
thus crystalline particles were present in the ``liquid'' side
and vice-versa, which confuse the structure near the interface.

\begin{figure}[htbp]
\includegraphics[scale=0.50]{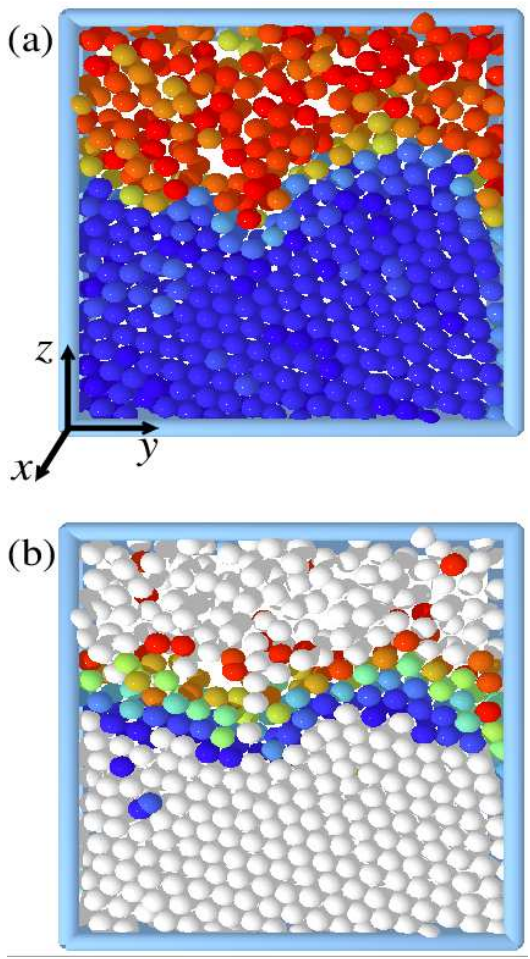}
\caption{\label{interface}
(Color)
(a) Picture of the liquid/crystal interface, showing a slice that is
two crystalline layers thick (4.8~$\mu$m).  Particles are colored
according to the number of ordered neighbors they have, $N_o$.
Blue particles have $N_o \geq 8$ and darker blue indicates
higher $N_o$.  Yellow, orange, and red particles have $N_o <
8$, with red particles having $N_o=0$ and the brightest yellow
particles having $N_o=7$.  The image has been slightly rotated
(by $3^{\circ}$ around an axis parallel to gravity) to view the
crystalline particles in-plane.  (b) Here the particles are colored
according to how much time they spend as ``crystalline'' particles
(with $N_o \geq 8$).  White particles spend all of their time as
crystalline or liquid-like, and the colored particles fluctuate
over the $\sim 1$~hr experiment.  Blue particles spend nearly all
their time as crystalline, light green corresponds to those which
spend half their time in each state, and red particles spend nearly
all their time as liquid-like.  While the coloring is based on the
time average of the data, the positions are considered at the same
time as shown in panel (a).
}
\end{figure}

In this work, we present confocal microscope observations of
an equilibrated colloidal crystal/liquid interface.  By following
the positions of several thousand colloidal particles on both
sides of the interface, we directly visualize the interface.
An example of our data is shown in Fig.~\ref{interface}(a), where blue
particles are crystalline and yellow/red particles are liquid-like.
This interface has a low surface tension and we see capillary waves.
We are able to remove the influence of these capillary waves from
the data and measure the intrinsic surface profile.  In particular,
we find that capillary waves cause an apparent broadening of the
surface, but the structure of the intrinsic surface is characterized
by a width of only $1.5d$ (in terms of the particle diameter
$d$).  This is the first direct experimental visualization of an
equilibrated interface.  A precise definition of the intrinsic
interface has not been made before, and we show that several
plausible definitions give slightly different results.

Our colloidal sample is described in the Materials and Methods
section.  Here, we briefly note that in our solvent the colloidal
particles have a slight charge.  Experimentally we observe that
the freezing transition volume fraction is at $\phi_{\rm freeze}
= 0.43$ and the melting transition is at $\phi_{\rm melt}= 0.49$,
to be compared with the hard sphere values of $\phi_{\rm freeze}
= 0.494$ and $\phi_{\rm melt}= 0.545$.  Our measured values are
similar to that seen for other experiments with similar colloidal
samples \cite{gasser01}.

\section{Results}

At each time step, we determine the crystalline region by using
the method of
bond-order parameters \cite{kegel06,gasser01,steinhardt83,wolde96}.
To do this, each particle $i$ at each time is characterized by a
normalized order parameter $\hat{\bf q}_{l}(i)$ with  $(2l+1)$
complex components
\begin{equation}
      \hat{q}_{lm}(i)=\frac{1}{\mathcal{N}B(i)}\sum_{j=1}^{B(i)}
      Y_{lm}(\hat{\bf r}_{ij}),
      \label{eq:qlm}
\end{equation}
where $\mathcal{N}$ is a normalization factor such that
$\sum_{m}\hat{q}_{lm}(i)\hat{q}_{lm}^{*}(i)=1$, $B(i)$ is the
number of neighbors of particle $i$, $\hat{r}_{ij}$ is the
unit vector pointing from particle $i$ to its $j$'th neighbor,
and $Y_{lm}$ is a spherical harmonic function.  Following prior
work, we use $l=6$ \cite{wolde96,gasser01}.  The neighbors of
a particle are defined as those with centers separated by less
than $1.41d$, which corresponds to the first minimum of the pair
correlation function $g(r)$ for the liquid region.  Two neighboring
particles are termed ``ordered neighbors'' if the complex inner
product $\sum_{m}\hat{q}_{lm}(i) \hat{q}_{lm}^{*}(j)$ exceeds a
threshold value of 0.5.  For each particle, the number of ordered
neighbors $N_o$ is determined.  Following the usual convention,
particles with $N_o \geq 8$ are classed as crystalline particles,
and the other particles are liquid-like particles \cite{wolde96}.
The advantages of using bond-order parameters is that they are
local measures of order, are somewhat insensitive to variations
in the number of neighbors each particle has,
and also do not depend on the specific
type of crystal \cite{steinhardt83,wolde96}.

Figure \ref{interface}(a) shows a snapshot of the sample, colored
according to the bond-order parameter method.
The crystalline side of the interface (blue particles) is
composed of hexagonal layers in random stacking (a mixture
of face-centered-cubic stacking [{\it abcabc...}] and
hexagonal-close-packed stacking [{\it ababab...}]).  This is
similar to the stacking seen in growing colloidal crystallites
\cite{gasser01}.  In particular, the data shown in this manuscript
are for a sample where 9 hexagonal layers are imaged, with stacking
{\it abacbcbac}.  Two other regions of this sample were imaged
and the results presented below do not vary in any significant way.

Typically, between two different phases, there will always exist
a surface energy defining the energetic cost of maintaining
the interface.  For fluid/fluid interfaces, this is also known
as the surface tension, although the concept is relevant for
solid/liquid interfaces as well (for example, in nucleation
of crystals \cite{gasser01}.)  For a hard sphere system, this
surface energy is entropic in origin, reflecting the difficulty
of packing particles at the interface.  Right at the interface,
particles can pack neither optimally for the crystal state, nor
optimally for the liquid state, thus resulting in an entropic
penalty that gives rise to a surface tension.  In our experiment,
the interface shown in Fig.~\ref{interface} is rough and varies in
time due to surface capillary waves.  Because the surface energy
is not extremely large compared to $k_B T$ (the thermal energy,
based on the temperature $T$ and Boltzmann's constant $k_B$),
thermal fluctuations allow these capillary waves to be observed in
our experiment.  The temporal variability of the sample is indicated
in Fig.~\ref{interface}(b), which colors only the particles which we
observe to spend some time in both crystalline and liquid structures
during the duration of our experiment (3750~s).  As can be seen,
the exact position of the interface thus fluctuates by as much as
$\sim 4$ particle diameters (in a ``peak-to-peak'' sense).

\begin{figure}[tbp]
\includegraphics[scale=0.40]{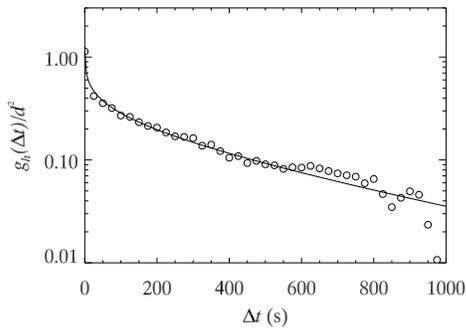}
\caption{\label{tension}
The circles show the temporal correlation function of interface
position, averaged over $x$ and $y$ (Eqn.~\protect\ref{cor}).
The line is a fit to Eqn.~\protect\ref{fit}, with 
interfacial stiffness
$\tilde{\gamma} = 0.93$~nN/m $= 1.2 k_B T / d^2$ and a capillary time
$\tau = 750$~s.  The small $k$ modes give rise to the very steep
drop of the correlation at short time scales, and the exact shape
of the theoretical curve for $\Delta t < 25$~s depends on how the
numeric integration of Eqn.~\protect\ref{fit} is done.
}
\end{figure}

To determine the surface tension, we follow the procedure of
Ref.~\cite{aarts04}.  Due to a slight purposeful density mismatch
(described in the Methods and Materials section), the
crystal/liquid interface is nearly parallel to the $xy$ plane; we
rotate the data by $15^\circ$ to align the interface with the
$xy$ plane.  Based on the new $z$ axis, 
at each time we locate the maximum $z$
position of crystalline particles as a function of $x$ and $y$
(coarse-grained in $x$ and $y$ over a distance $1.035d$,
the spacing between crystalline layers, where $d$ is the diameter
of an individual particle).  This gives us the
interface height $h(x,y,t)$.  While our data is only over a short
spatial extent in $x$ and $y$ ($9d \times 24d$),
we have good temporal data ($0
\leq t \leq 3750$~s) and thus we calculate the temporal dynamical
correlation function:
\begin{equation}
\label{cor}
g_h(\Delta t) = \langle [h'(x,y,t)][h'(x,y,t+\Delta t)]
\rangle_{x,y,t}
\end{equation}
where $h'(x,y,t) = h(x,y,t) - \langle h(x,y,t) \rangle_t$ represents
the fluctuations of the interface about its time-averaged position,
and the angle brackets $\langle ... \rangle_{x,y,t}$ indicate an
average over space and time.  We plot this dynamic correlation
function in Fig.~\ref{tension} (circles).  The intercept is at
$g_h(\Delta t=0) \approx 1.1 d^2$; thus the root-mean-square
width of the capillary wave fluctuations is approximately $d$,
as compared to the slightly larger amplitude of the peak-to-peak
fluctuations seen in Fig.~\ref{interface}(b).

The capillary waves are limited by the interfacial stiffness
$\tilde{\gamma}$, rather than the surface tension $\gamma$
\cite{hoyt01}.
For a crystal / liquid interface, $\gamma$ is usually anisotropic
and depends on the crystal orientation.  Fluctuations in the
surface thus depend both on the surface tension $\gamma$ (related
to interfacial stretching) as
well as second derivatives of $\gamma$ with respect to angles
away from the interface normal (related to interfacial bending)
\cite{laird05b,hoyt01,du07}.
The orientational average of
$\tilde{\gamma}$ is the bulk surface tension $\gamma$, but
measuring
this requires vastly more data than we have
\cite{laird05b,hoyt01,du07}.  To extract
the interfacial stiffness $\tilde{\gamma}$, we fit $g_h(\Delta t)$
using the results of capillary wave theory \cite{aarts04,steyerl98}.
Overdamped capillary waves with wave number $k$ should decay as
\begin{equation}
\label{damping}
\exp[-t(\tilde{\gamma} k + g \Delta \rho_i/k)/(2 \eta)],
\end{equation}
with gravitational acceleration $g$, density difference
$\Delta \rho_i$ (across the interface), and viscosity $\eta$
(equal to the sum of the
viscosities of the two phases) \cite{steyerl98}.  By
equipartition, the amplitude of the Fourier component $h_k$ of
the interface displacement contributes as
\begin{equation}
\label{amplitude}
\langle | h_k |^2 \rangle = \frac{k_B T}{\tilde{\gamma} L^2}
\frac{1}{k^2 + \xi^{-2}}
\end{equation}
with $L$ the lateral system size, and using the capillary length
$\xi = \sqrt{\tilde{\gamma} / (g \Delta \rho_i)}$ \cite{penfold01}.
As in Ref.~\cite{aarts04}, we define the nondimensional
$\bar{k} = \xi k$, and then combine these two results to
calculate the theoretical dynamic
correlation function as an integral
\begin{equation}
\label{fit}
g_h(\Delta t) = \frac{k_B T}{2 \pi \tilde{\gamma}}
\int^\infty_0 d\bar{k} \bar{k}
\frac{\exp[-(\bar{k} + \bar{k}^{-1}) \Delta t/(2\tau)]}{1 +
\bar{k}^2}
\end{equation}
with $\tau = \xi \eta / \tilde{\gamma}$ as the capillary time, the
characteristic time scale for decay of interfacial fluctuations
\cite{penfold01}.  In particular, $\tau$ corresponds to the time
scale for the slowest decaying wavelength, which is at the capillary
length scale $\xi$.

We vary $\tau$ and $\tilde{\gamma}$ to find the best fit to
our data, and plot this as the solid line in Fig.~\ref{tension},
finding excellent agreement using $\tilde{\gamma} = 1.2 k_B T/d^2$.
Our value of $\tilde{\gamma}=1.20 \pm 0.05$ (in units of $k_B
T/d^2$) is a similar order of magnitude to previously found
values of $\gamma$ for hard spheres, which range from 0.11 to 0.78
\cite{laird05b,laird05,gasser01,vanmegen97,auer01,song06}.  Our
measured $\tilde{\gamma}$ is larger than these values of $\gamma$,
perhaps due to the above-noted anisotropy of $\tilde{\gamma}$,
for which we may have a stiff direction \cite{laird05b}.


We note several limitations with our measurement of the
interfacial stiffness.  First, our result is calculated for
one particular interface orientation, rather than averaging
over many different crystal orientations.  Second, the theory of
Refs.~\cite{aarts04,penfold01} is derived for liquid-gas interfaces,
where the surface tension is isotropic.  As noted above, the local
interfacial stiffness is an adequate replacement for the isotropic
surface tension \cite{laird05b}.  Third, the theory also assumes
that both phases have a well-defined viscosity, whereas one of
our phases is a colloidal crystal.  The crystalline phase could
be considered as a liquid with a very large viscosity, but this
would then predict a very large capillary time, in contrast with
our observed $\tau = 750$~s.  This value is consistent with a
viscosity $\eta=50$~mPa$\cdot$s$\approx 20 \eta_0$ (compared with
the solvent viscosity $\eta_0$).  This is approximately half of
what could be expected for high volume fraction colloidal samples
at volume fractions $\phi = 0.43 - 0.49$ \cite{cheng02}, given that
$\eta$ here represents the sum of the viscosities of the two phases.
To summarize, our crystal-liquid interface has capillary waves
which fluctuate in a way consistent with predictions for liquid-gas
capillary waves, but interpretations of this observation should
be done cautiously.

Additionally we note that the space-averaged height $\langle
h(x,y,t) \rangle_{x,y}$ fluctuates but does not monotonically
increase or decrease, confirming that we observe an equilibrium
interface rather than a system which is crystallizing or melting.
However, as can be seen in Fig.~\ref{interface}(b), the exact
interface position fluctuates within a certain range over the
duration of our experiment, precluding us from an ensemble-averaged
sort of measurement.  For a longer duration experiment, the
time-averaged shaded region in Fig.~\ref{interface}(b) would
presumably meander less and be more uniform across the $y$
direction.

\begin{figure}[tbp]
\includegraphics[scale=0.65]{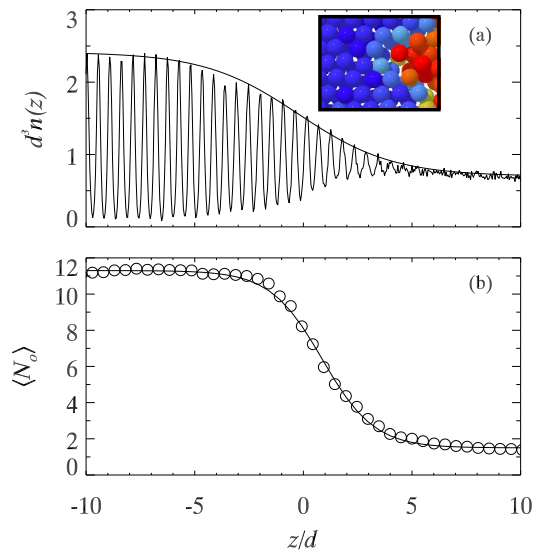}
\caption{\label{profiles} 
Two views of the interfacial profile in the ``laboratory''
reference frame.
(a) Interfacial profile, showing the
number density profile $d^3 n(z)$ as a function of
$z/d$.  The smooth curve is a hyperbolic tangent fit to the
envelope, with 10-90 width of $9.6d$.  The inset shows the
crystal orientation used to determine this curve,
after the sample has been rotated slightly (see text
for details).  
(b) Bond-order profile,
showing the average number of ordered neighbors particles have,
as a function of $z/d$.  The dotted curve is again a hyperbolic
tangent fit, with a 10-90 width of $6.08d$.  By definition, $\langle
N_0 \rangle = 8$ at $z=0$.  For both (a) and (b), the data are
time averaged over 2000~s ($80$ 3D images).
}
\end{figure}

We now turn to the structural details of the crystal-liquid
interface.  As mentioned above, we rotate the data around the
$x$ axis by 15$^\circ$ clockwise (as seen from the view in
Fig.~\ref{interface}), so that the crystal structure is aligned
with the new $z$ axis; see inset of Fig.~\ref{profiles}(a).
We then plot the number density $d^3 n(z)$ as a function of $z/d$
in Fig.~\ref{profiles}(a), and see oscillations similar to what
has been seen in experiments \cite{kegel06} and simulations
\cite{laird98,broughton86,huitema99}.  The smooth line in
Fig.~\ref{profiles}(a) is a hyperbolic tangent fit to the data.
Following previous work, we define the interfacial width as the
10-90 width, where the hyperbolic tangent function goes from 10\% of
its value to 90\% of its value.  We find $W_{10-90} = 9.6d$, quite
similar to what has been seen previously \cite{kegel06,laird98}.
Some of this width is due to capillary waves, and some may be due to
particles in the crystalline lattice (but near the interface) having
larger fluctuations around their mean positions \cite{laird98}.
We checked for this latter influence by low-pass filtering
\cite{laird98} the density profile of Fig.~\ref{profiles}(a)
and found a similar $W_{10-90}$ width, although it depends on the
details of the filter; this filtering also shifts the halfway point
of the hyperbolic tangent fit toward the liquid side by $4d$.
We also calculate the width based on the structural ordering,
by plotting the average number of ordered neighbors $\langle
N_o \rangle$ as a function of $z'/d$ in Fig.~\ref{profiles}(b).
Based on this measure, the profile is slightly sharper, with
$W_{10-90} = 6.1d$; this sharpness agrees with previous results
\cite{laird98,kegel06}.

\begin{figure}[tbp]
\begin{center}
\includegraphics[scale=0.40]{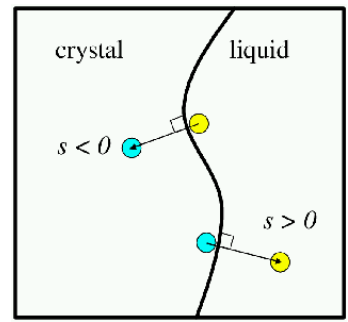}
\end{center}
\caption{\label{sketch} 
Sketch defining the distance from the interface $s$.
For crystalline particles, the distance $s$ is found
to the closest non-crystalline particle (any particle with
$N_o<8$).  This distance is defined to be negative. 
For liquid particles, the distance $s$ is the closest distance to
the nearest crystalline particle ($N_o \geq 8$).  Based on this
definition, the region $-d < s < d$ is excluded, as no two particles
are closer than their diameters $d$.  Thus, we shift the
distance $s$ by $+d/2$ for crystalline particles, and by $-d/2$
for liquid particles.  This leaves a gap of size $d$, reflecting
that the interface truly lies between the
crystalline and liquid particles.
}
\end{figure}

However, Fig.~\ref{interface} implies that measurements that average
over $x$ and $y$ will artificially broaden the interface, given
that the interface is not flat.  Due to the fractal structure of
the capillary waves, the interfacial width depends on the total
system size.  Our measured widths depend on the size of the
observation region, and would increase if the observed region
was larger and encompassed more of the sample \cite{chacon06}.
Our data allow us a closer examination of the detailed behavior
near the interface.  For each particle, we calculate the distance
$s$ to the instantaneous interface position using the procedure
illustrated in Fig.~\ref{sketch}.

\begin{figure}[tbp]
\includegraphics[scale=0.64]{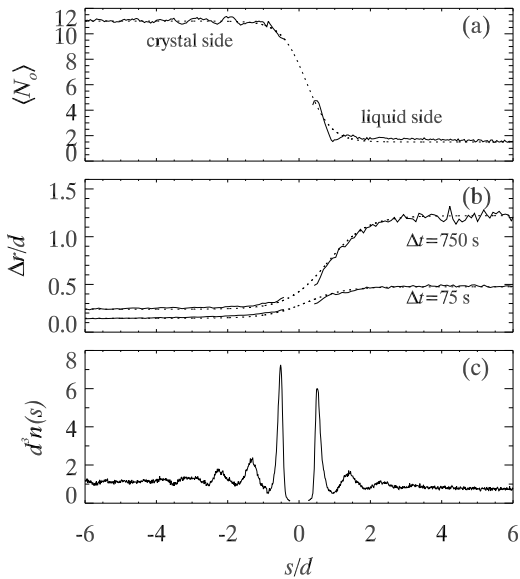}
\caption{\label{newprof} 
Three views of the intrinsic interface profile.
(a) Bond-order profile (solid line), showing the average number of ordered
neighbors of particles as a function of $s/d$.  $s$
is the distance from the interface, as described in the text and
Fig.~\protect\ref{sketch}.  The dotted line is a hyperbolic
tangent fit with $W_{10-90}=1.5d$.
(b) Average distance $\Delta r$
moved in a given $\Delta t$ as indicated, as a function of $s/d$.
The dotted lines are hyperbolic tangent fits, each with
$W_{10-90}=2.4d$.  Here $s$ is the distance measured at the
particle's initial position at the start of the displacement
$\Delta r$.
(c) Nondimensional number density $d^3 n$ as a function of
$s/d$.  
}
\end{figure}

To examine the transition from order to disorder, we calculate
the average number of ordered neighbors $\langle N_o \rangle$ now
as a function of $s$, and plot the result in
Fig.~\ref{newprof}(a).  By definition, $N_o \geq 8$ for
crystalline particles and $N_o \leq 7$ for liquid-like particles,
thus producing a gap in $\langle N_o \rangle$ at the interface.
Figure~\ref{newprof}(a) shows that based on the ordering, the
interface is quite sharp, with a width $W_{10-90}= 1.5 d$, in contrast
to the interpretation of Fig.~\ref{profiles}.  By this measure,
the interface is locally a sharp, well-defined interface, as has
been conjectured \cite{stillinger65,muller05}.

Figure \ref{newprof}(b) shows the mobility $\Delta r/d$,
measured across the interface (defined for two different time
scales $\Delta t$).  Here we see the interfacial width is
$W_{10-90}=2.4d$, significantly broader than that determined by
structure but still fairly narrow, supporting the interpretation
of a meandering but sharp interface.  On the liquid side, the
mobility is still increasing for $s>d$ where $\langle N_o \rangle$
has reached the steady-state liquid value; the halfway point of
the hyperbolic tangent fit is at $s=0.2d$ for the 
$\Delta t=75$~s data and at $s=0.8d$ 
for the 
$\Delta t = 750$~s data.  This result
is understandable as the crystalline particles form a relatively
immobile region:  thus the more mobile
liquid-like particles are
slowed by proximity to the crystal, similar to a hydrodynamic
``no-slip'' condition.  Examining the individual components of
motion ($\Delta x, \Delta y, \Delta z$) we find no significant
differences, indicating that the motion is isotropic, as seen in
simulations \cite{laird98}.  While our data for $\Delta t > 750$~s
are noisy, we can estimate that the diffusivity on the liquid
side is approximately 50 times smaller than the diffusivity in a dilute
suspension, in reasonable agreement with hard sphere simulations 
\cite{laird98}.

To examine the intrinsic density profile, we plot the 
number density $d^3 n(s)$ in Fig.~\ref{newprof}(c).  The crystal
structure has a fixed orientation, and thus relative to the curving
interface, the layers seen in Fig.~\ref{profiles}(a) diminish
away from the interface (the region $s<0$).  On the
liquid side ($s>0$), layering relative to the intrinsic
interface is seen with two clear layers, fewer layers than 
the crystal side.  This reinforces that the layering shown in
Fig.~\ref{profiles}(a) for the region $z>0$ is mostly due to the
uneven interface.

\section{Discussion}

Note that in this work we have studied slightly charged
particles rather than ideal hard spheres.  Perhaps surprisingly,
our results compare reasonably well with hard sphere simulations
such as Refs.~\cite{laird98,laird05,laird05b,song06}.  Thus the
influence of the charges shift the phase boundaries as noted
above (a freezing volume fraction $\phi_{\rm freeze}=0.49$ for
our experiments as compared to 0.545 for hard spheres), but the
charges do not seem to strongly modify the structure or dynamics
of the interface.  It is likely that particle motion is
somewhat slowed due to smaller effective cage sizes \cite{weeks02}
and thus modifying the plateau levels of Fig.~\ref{newprof}(b).
However, this does not seem to modify the extent of the
transition region of the interface, and indeed one might expect
that a long-range interaction would only broaden the interface,
thus making it more striking that we observe such narrow
interface widths.

We have studied the equilibrium crystal/liquid interface of a dense
colloidal suspension.  We observe a rough interface with width
6-10$d$ depending on the property examined.  Capillary waves account
for some of this broadening, although overall the mean square
fluctuations from capillary waves have magnitude $g_h(\Delta t=0)
\approx 1.1 d^2$.  Thus, there may be a base structure comprised
of crystalline facets, as suggested by the darkest particles
shown in Fig.~\ref{interface}(b).  These facets may be slower
to rearrange, and their potential existence likely broadens the
observed interface width.

We also studied the profile in the direction locally perpendicular
to the rough interface, and find that this intrinsic interface is
much sharper, 1.5$d$ and 2.4$d$ based on structure and dynamics,
respectively.  These results emphasize that the observed
broadness of the interface is only apparent, and depends on the
measurement length scales \cite{chacon06} and time
scales \cite{laird98}.  The underlying spatial transition from one
phase to the other is quite sharp, confirming the classic picture
\cite{stillinger65,muller05}.  Our data provide a useful test
for future models of the intrinsic profile.

\section{Materials and Methods}

Our samples are composed of colloidal poly-methyl(methacrylate)
particles, sterically stabilized to prevent aggregation
\cite{pusey86}.  The particles have mean diameter $d=2.30 \pm
0.02$~$\mu$m and a polydispersity of approximately 5\%.  The solvent
is a mixture of cyclohexylbromide and decalin, chosen to closely
match the density and index of refraction of the particles, with
viscosity $\eta_0 = 2.25$~mPa$\cdot$s \cite{dinsmore01}.  However,
we intentionally slightly density mis-match the fluid by adding
an excess of decalin (which is more dense than the particles).
This lets us gravitationally induce a crystal-liquid transition
within our sample chamber.  Our particle/fluid density mismatch is
less than $\Delta \rho \approx 5 \times 10^{-4}$~g/ml, giving us a
gravitational scale height $k_B T / \Delta \rho V g > 130 \mu$m,
using the temperature $T$ (295 K
for these experiments), the particle volume $V = \frac{\pi}{6}
d^3$, and the gravitational acceleration $g$.  In our solvent, the
particles have a slight charge, shifting the freezing transition
volume fractions to $\phi_{\rm freeze} = 0.43$ and the melting
transition to $\phi_{\rm melt}= 0.49$, similar to values for other
weakly charged colloidal PMMA experiments \cite{gasser01}.
The existence of the charge is also apparent from comparisons of
the pair correlation function $g(r)$ taken at different volume
fractions.  For hard spheres, the first peak of $g(r)$ should
always be at $r=d$, whereas for our samples we find the peak at
$r>d$ and the peak position varies slightly with $\phi$.
However, this peak position is still fairly close to $d$; for
example, in the crystalline region we find the peak at $r=1.04d$
and in the liquid region we find the peak at $r=1.15d$,
suggesting that the charging is not too extreme \cite{yethiraj03,royall06}.

Our microscope sample chambers have dimensions 0.4 mm $\times$ 3
mm $\times$ 30 mm, and are stored with the long dimension oriented
vertically.  Because of the slight density mismatch between the
solvent and the particles, gravity sets up a slight concentration gradient.
We let the samples equilibrate in the sample chamber for more than
one month before taking data.  We make one important modification
to our microscope: the microscope base is tilted 90$^\circ$ so that
the objective lens points horizontally \cite{koehler04}.  The sample
is placed on the microscope stage with the orientation kept the same
as the storage conditions.  Thus we study a completely equilibrated
sample at a stable crystal-liquid interface, and the direction of
gravity ($-\hat{z}$) points perpendicular to the optical axis.

The particles are dyed with rhodamine 6G, so that they
can be viewed with a laser scanning confocal microscope
\cite{dinsmore01,prasad07}.  We acquire images of size $20 \times
55 \times 60$~$\mu$m$^3$ once every 25~s, where the long direction
($z$) is parallel to gravity.  The images were taken from 20 to
40 $\mu$m away from the coverslip, to avoid direct influence of
the walls.  As noted above, the full sample size is $0.4 \times
3 \times 30$~mm$^3$, where the smallest dimension is $x$ and the
largest direction is $z$; thus the imaged region is only a small
volume within the sample.  The 3D images are analyzed to determine
particle positions with a resolution of 0.1~$\mu$m parallel to the
optical axis ($x$) \cite{dinsmore01,crocker96}.  and a resolution
of 0.05~$\mu$m perpendicular to the optical axis ($y$ and $z$)
Because this is a dense sample, particles do not move far between
images, and thus we follow their motion using standard particle
tracking techniques \cite{crocker96}.

\section{Acknowledgments}

We thank D.~G.~A.~L.~Aarts, M.~Asta, G.~C.~Cianci, B.~B.~Laird,
W.~K.~Kegel, and T.~Witten for helpful discussions.  This material
is based upon work supported by the National Science Foundation
under Grant No.~0239109.

\bibliography{weeks}

\end{document}